\documentclass{INTERSPEECH2023}
\usepackage{amsmath,graphicx,epstopdf,booktabs,multicol,multirow,enumitem,footnote,type1cm}
\usepackage[table]{xcolor}
\usepackage{color,CJKutf8}

\interspeechcameraready

\title{ZeroPrompt: Streaming Acoustic Encoders are Zero-Shot Masked LMs\vspace{-8pt}}
\name{Xingchen Song$^{1,2,3}$,~Di Wu$^{2,3}$,\\~Binbin Zhang$^{2,3}$,~Zhendong Peng$^{2,3}$,~Bo Dang$^{3}$,~Fuping Pan$^3$,~Zhiyong Wu$^1$\vspace{-6pt}}
\address{$^1$Tsinghua Univ., Beijing, China $^2$Horizon Inc., Beijing, China $^3$WeNet Open Source Community}
\email{xingchen.song@horizon.ai}

\begin{document}

\maketitle

\begin{abstract}
In this paper, we present ZeroPrompt (Figure 1-(a)) and the corresponding Prompt-and-Refine strategy (Figure 3), two simple but effective \textbf{training-free} methods to decrease the Token Display Time (TDT) of streaming ASR models \textbf{without any accuracy loss}. The core idea of ZeroPrompt is to append zeroed content to each chunk during inference, which acts like a prompt to encourage the model to predict future tokens even before they were spoken. We argue that streaming acoustic encoders naturally have the modeling ability of Masked Language Models
and our experiments demonstrate that ZeroPrompt is engineering cheap and can be applied to streaming acoustic encoders on any dataset without any accuracy loss. Specifically, compared with our baseline models, we achieve 350 $\sim$ 700ms reduction on First Token Display Time (TDT-F) and 100 $\sim$ 400ms reduction on Last Token Display Time (TDT-L), with theoretically and experimentally equal WER on both Aishell-1 and Librispeech datasets.
\end{abstract}

\section{Introduction}
In the past few years, end-to-end models, such as connectionist temporal classification (CTC)~\cite{ctc}, RNN-Transducer (RNN-T)~\cite{rnnt}, and attention-based encoder-decoder (AED)~\cite{speech-transformer} models, have achieved significant success on various ASR tasks. Recently, there has been a growing interest in developing end-to-end ASR models with streaming capability. Among them, chunk-based acoustic encoders~\cite{dual,saa,u2} have gained popularity and have been adopted in many previous works. These methods utilize bi-directional recurrent networks~\cite{blstm} or fully-connected self-attention networks~\cite{transformer} within a chunk. In this work, we primarily focus on chunk-based methods due to their full-context utilization in a chunk.

\begin{figure}[!h]
	\centering
	\vspace{-10pt}
	\includegraphics[scale=0.45]{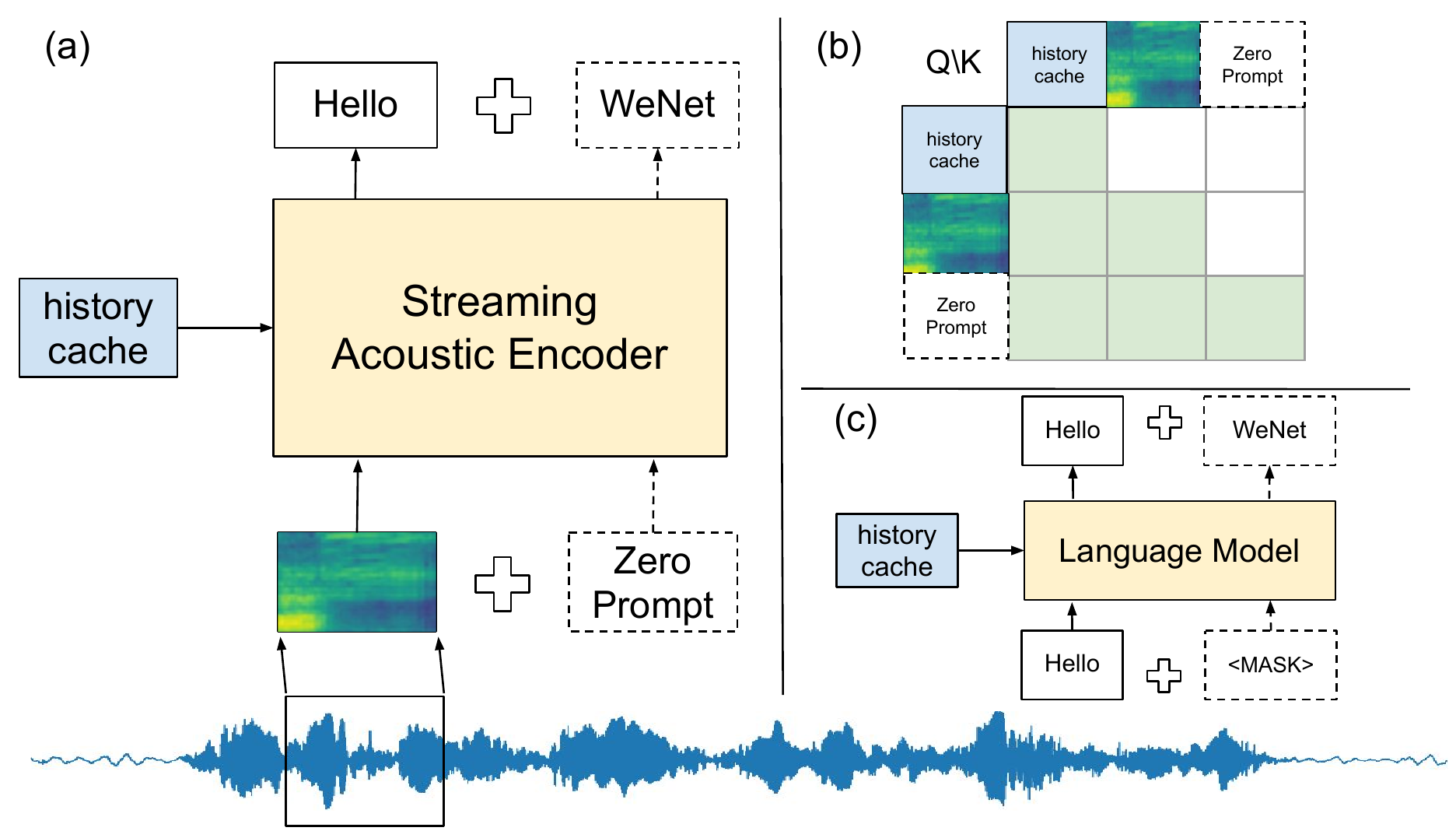}
	\vspace{-10pt}
	\caption{(a) Illustration of ZeroPrompt. (b) To keep the prediction of the current chunk not affected by zeroed future frames, we use a chunk-level autoregressive attention mask. (c) A symmetrical perspective on Masked LM.}
	\vspace{-10pt}
	\label{fig:zeroprompt}
\end{figure}

In streaming scenarios such as real-time subtitles, ASR systems need to decode speech with low latency, producing words as soon as possible~\cite{upl}. A straightforward way to reduce latency is directly decreasing chunk size (i.e., from 640ms to 320ms). However, there is often a trade-off between performance and latency and lower chunk size usually leads to higher WER. Another way to reduce latency is to apply regularization either on loss function~\cite{fastemit,peak-first-ctc} or input spectrogram~\cite{trimtail} to push forward the emission of tokens. While being successful in terms of reducing the token emission latency of streaming ASR models, the definition of token emission latency (i.e., The timestamp or frame index when the model predicts the token) underestimates the true user-perceived latency (such as Token Display Time) in chunk-based models, since they do not account for chunk cumulative time (a.k.a, the time to wait before the input signal forms a chunk). Here, we further provide an example to explain why token emission latency does not correlate well with our notion of user-perceived latency. In Figure 2, assume the second char of the recognition result happens at 1000ms and is pushed forward to 800ms after training with emission regularization, the model still needs to wait until 1200ms to form a valid chunk and hence start to decode and emit the second char.

To better measure the latency terms that accurately capture the user-perceived latency, we propose two metrics as illustrated in Figure 2: \textbf{\textit{First Token Display Time} (TDT-F)} and \textbf{\textit{Last Token Display Time} (TDT-L)} - the minimum chunk cumulative time required to output the first or last character. In real-time subtitle scenarios, those metrics can be used to evaluate the initial on-screen time of the first and last characters. For simplicity, we ignore the chunk computation time because it is usually much smaller than the chunk cumulative time, i.e., inference one chunk with 640ms chunk size usually takes only 50ms on a desktop CPU using single thread.

\begin{figure*}
	\centering
	\vspace{-10pt}
	\includegraphics[scale=0.5]{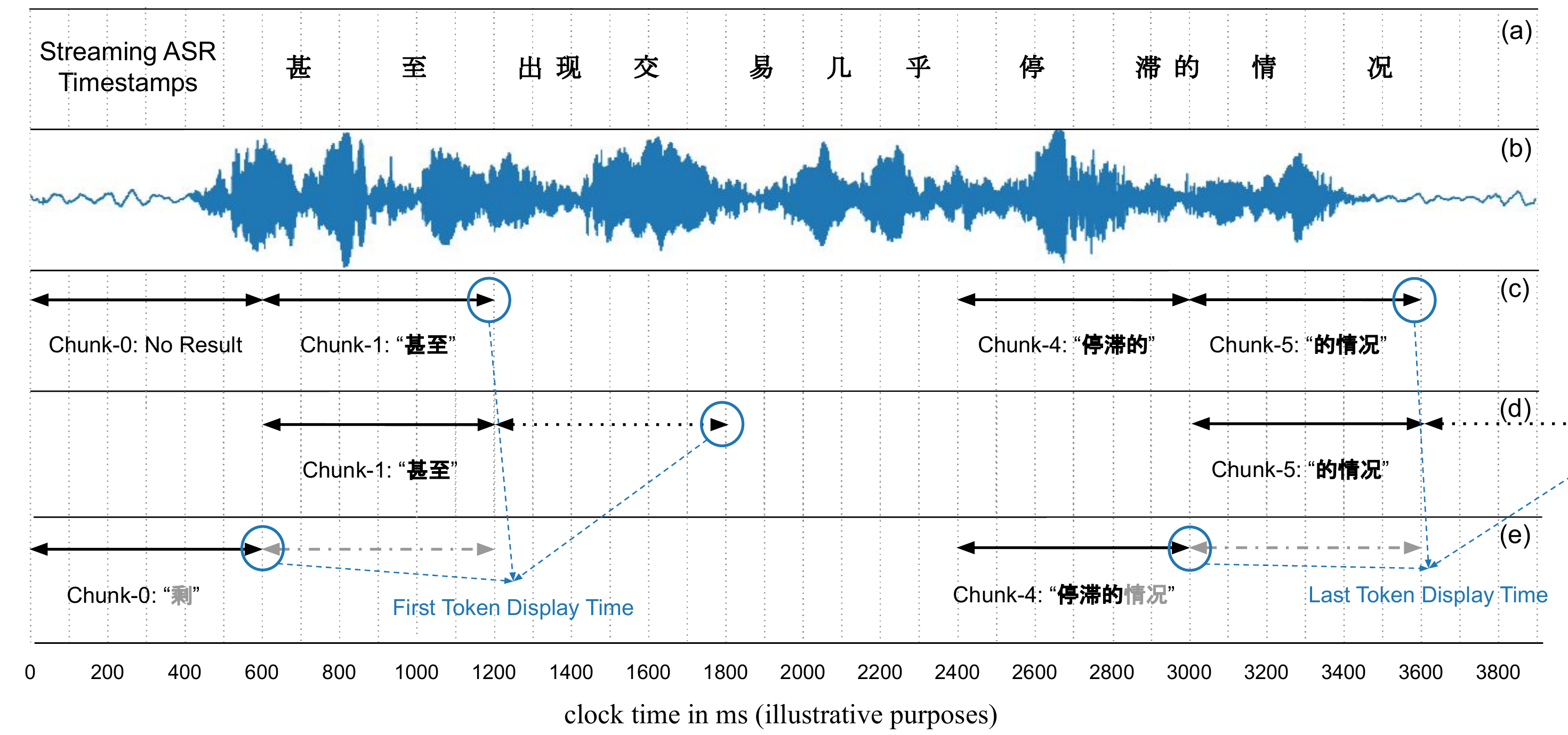}
	\vspace{-5pt}
	\caption{Illustration of timeline and latency metrics of a streaming ASR system. From top to bottom: (a) Streaming ASR timestamps. (b) Waveforms. (c) Causal method, 600ms chunk size without right context. (d) LookAhead methods~\cite{lookahead1,lookahead2}, 600ms chunk size with 600ms \textbf{real} right context (dotted line in black, a.k.a. LookAhead chunk). (e) ZeroPrompt method, 600ms chunk size with 600ms zeroed context (dash-dotted line in grey, a.k.a ZeroPrompt chunk), the black tokens mean predictions from the current chunk while grey tokens mean predictions from ZeroPrompt chunk.}
	\vspace{-15pt}
	\label{fig:latency}
\end{figure*}

In this paper, we explore a training-free method, called ZeroPrompt, which appends zeroed content to each chunk to prompt the model to predict future tokens through its zero-shot ability of Masked LMs that has been implicitly learned during training. We argue that previous works mainly focus on the \textbf{decoder part} of encoder-decoder E2E ASR structure rather than the \textbf{encoder part} to estimate the internal LM because the encoder part is usually optimized with CTC loss and CTC is generally not considered capable of modeling context between output tokens due to conditional independence assumption~\cite{adapter}. However, CTC-optimized ASR encoders learn the training data distribution and are affected by the frequency of words in the training data. The CTC-optimized encoder therefore at least has the modeling ability of a unigram LM to do something like MaskPredict (see Figure 1-(a) and Figure 1-(c) for a clearer comparison between ZeroPrompt and MaskPredict~\cite{bert}), and this paper aims to adopt this zero-shot ability to predict future tokens even before they were spoken and hence greatly reduce the TDT-F \& TDT-L during inference. Besides, to ensure that the final decoding result (or WER) is not affected, we propose to use a chunk-level autoregressive attention mask described in Figure 1-(b), coupled with a revision strategy called Prompt-and-Refine, to iteratively predict future tokens and refine them when the real future chunk arrives (see Figure 3 for a detailed example). Experimental results in Section 3 demonstrate that our methods have many advantages which can be summarized as:
\begin{itemize}
  \item ZeroPrompt does not require any model re-training and it takes nearly zero engineering cost to plugin any chunk-based streaming decoding procedure.
  \item ZeroPrompt can not only decrease the TDT-F \& TDT-L for \textbf{partial} recognition results but also keep the WER unaffected for \textbf{final} recoginition results. In other words, we achieve the theoretically and experimentally best trade-off between latency and WER.
\end{itemize}

\section{Proposed Methods \& Related Works}
As shown in Figure 1-(a), during inference, we process the utterance chunk-by-chunk, and append a certain number of zeroed future frames (called ZeroPrompt chunk) to each chunk. The history cache, current chunk, and ZeroPrompt chunk are together fed to the acoustic encoder to produce the prediction for both the current chunk (``Hello'') and the ZeroPrompt chunk (``WeNet''). Figure 1-(c) reveals that streaming acoustic encoders are zero-shot Masked Language Models (Masked LMs) and hence the ability of ZeroPrompt is something like MaskPredict used in standard Masked LMs.

This paper is related to LookAhead methods which use either \textbf{real}~\cite{lookahead1,lookahead2,revision} or \textbf{fake}~\cite{cuside} future frames. In previous work~\cite{lookahead1,lookahead2}, using the real right context requires waiting for the arrival of future content, which results in additional latency (Figure 2-(d)). Another study~\cite{revision} proposed a 2-pass strategy to process the current chunk first and revise it later once the future chunk is received, but its TDT-F \& TDT-L are identical to our baseline causal method (Figure 2-(c)) when compared within equal chunk size.

To avoid waiting for future context, CUSIDE~\cite{cuside} proposed an extra simulation encoder that is jointly trained with the ASR model and optimized with a self-supervised loss called autoregressive predictive coding (APC)~\cite{apc} to simulate a certain number of future frames for every chunk. While both CUSIDE and ZeroPrompt generate fake future information to avoid waiting time, they differ in how they utilize the generated futures. Specifically, ZeroPrompt directly concatenates the decoding results from the current chunk (black tokens in Figure 2-(e)) and ZeroPrompt chunk (grey tokens in Figure 2-(e)), whereas CUSIDE only uses the result from the current chunk (black tokens in Figure 2-(d)) as decoding output, and the simulated future is only used to enhance the recognition accuracy of the current chunk. Due to the different usage of the fake future content, the TDT-F \& TDT-L of CUSIDE are still identical to our causal baseline under equal chunk size. Moreover, ZeroPrompt uses much simpler zero padding to generate fake futures, so it does not require any extra parameters or model re-training compared to CUSIDE. Thanks to the internal ability of the Masked LM that is implicitly learned by the streaming encoder during training, ZeroPrompt can emit certain tokens even if the input is all zero.

\begin{figure}[!h]
	\centering
	\vspace{-8pt}
	\includegraphics[scale=0.45]{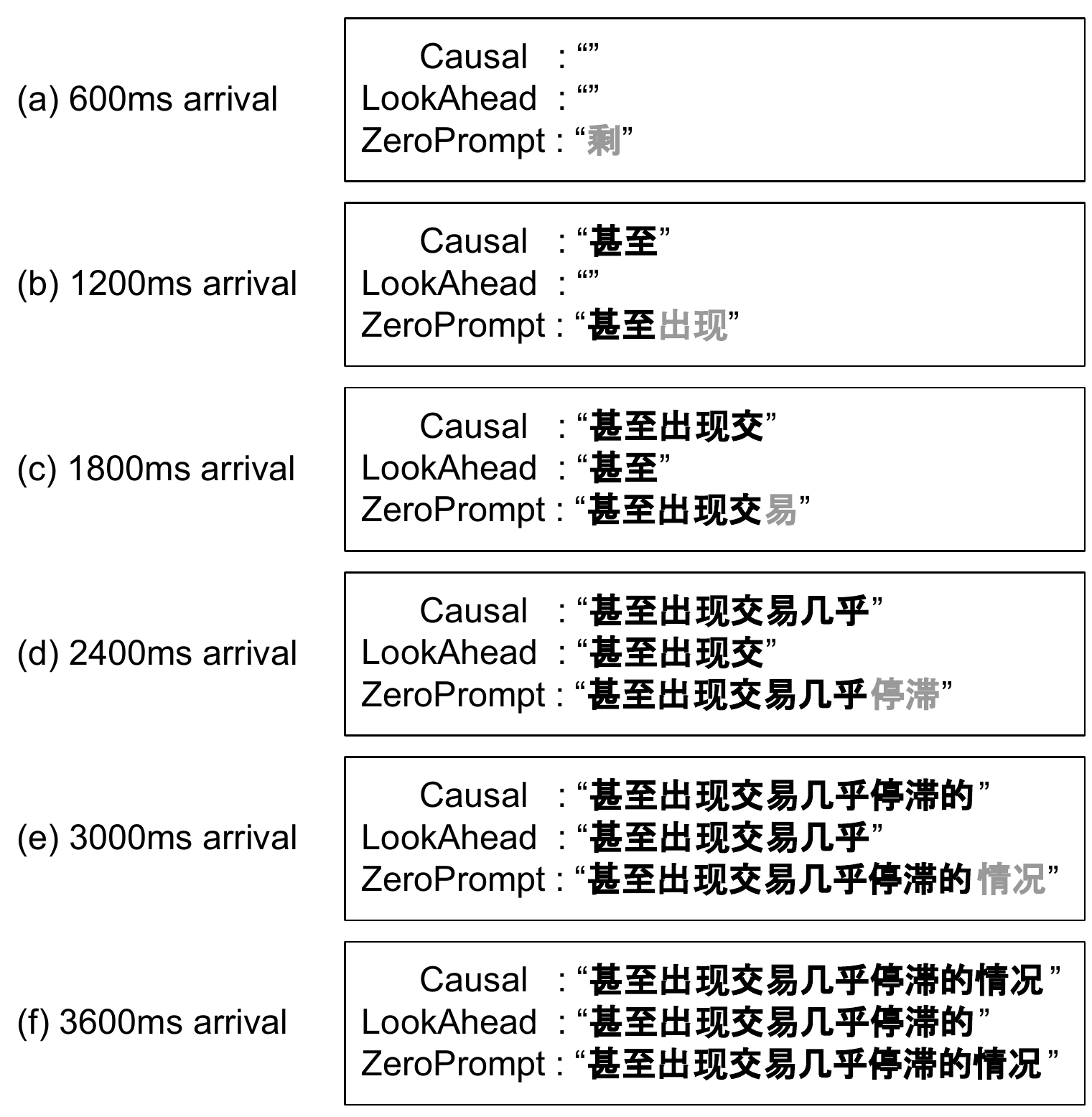}
	\vspace{-10pt}
	\caption{Comparison of on-screen time among three methods. We can clearly see that ZeroPrompt significantly improved the user-perceived latency. By comparing the result of ZeroPrompt in (a) \& (b), we observe that the mistake made by the first ZeroPrompt chunk is quickly fixed after the arrival of the second chunk which contains the real infos of the first few characters, this is so called Prompt-and-Refine.}
	\vspace{-8pt}
	\label{fig:output}
\end{figure}

\begin{table*}[!htp]
	\centering
	\caption{Comparison of different ZeroPrompt length across different chunk size on different dataset. From left to right: (a) length of ZeroPrompt. (b) First Token Display Time (TDT-F). (c) Last Token Display Time (TDT-L). (d) Prompts Error Rate for First chunk (PER-F). (e) Prompts Error Rate for Last chunk (PER-L). (f) Prompts Error Rate for All chunks (PER-A). (g) Word Error Rate (WER, \textit{1st-pass Greedy Search / 2nd-pass Rescore}). (h) Real Time Factor (RTF, \textit{1st-pass Greedy Search / 2nd-pass Rescore}, tested on Intel(R) Core(TM) i5-8400 CPU @ 2.80GHz using int8 quantization and single-thread). (i) Prompts Per Chunk (PPC). We note that the PER of Librispeech is significantly lower than that of Aishell-1. This is because we decode Librispeech using Byte Pair Encoding (BPE) but calculate the Prompts Error Rate using English characters. A BPE usually consists of several characters, and even if the BPE is incorrect, there may be correct letters, in other words, the denominator of PER increases while the numerator decreases.}
	\scalebox{0.73}{
		\begin{tabular}{c|cc|ccc|ccc}
			\toprule[1.5pt]
			(a) ZeroPrompt & (b)TDT-F & (c) TDT-L & (d) PER-F (\%) & (e) PER-L (\%) & (f) PER-A (\%) & (g) WER (\%) & (h) RTF & (i) PPC\\
			\midrule
			\multicolumn{9}{c}{Aishell-1 (test), 104765 total characters, 7176 total sentences} \\
			\midrule
			\multicolumn{9}{c}{640ms chunk size with 59081 total chunks} \\
			\midrule
			0ms & 1279ms \textcolor{green}{($\sim$)} & 4806ms \textcolor{green}{($\sim$)} & - & - & - & 5.81 / 5.05 & 0.04351 / 0.05063 & - \\
			80ms & 1272ms \textcolor{green}{($\downarrow$7)} & 4762ms \textcolor{green}{($\downarrow$44)} & 87 / 2191 = 3.9\% & 7 / 947 = 0.7\% & 442 / 12059 = 3.6\% & 5.81 / 5.05 & 0.04816 / 0.05495 & 0.20 \\
			160ms & 1234ms \textcolor{green}{($\downarrow$45)} & 4706ms \textcolor{green}{($\downarrow$100)} & 266 / 4351 = 6.1\% & 19 / 1937 = 0.9\% & 1162 / 23450 = 4.9\% & 5.81 / 5.05 & 0.05009 / 0.05722 & 0.39 \\
			320ms & 876ms \textcolor{green}{($\downarrow$403)} & 4603ms \textcolor{green}{($\downarrow$203)} & 1834 / 7457 = 24.5\% & 152 / 4211 = 3.6\% & 5183 / 46180 = 11.2\% & 5.81 / 5.05 & 0.05378 / 0.06282 & 0.78 \\
			640ms & 646ms \textcolor{green}{($\downarrow$633)} & 4472ms \textcolor{green}{($\downarrow$334)} & 5816 / 10150 = 57.3\% & 867 / 7408 = 11.7\% & 20181 / 71091 = 28.3\% & 5.81 / 5.05 & 0.06447 / 0.07425  & 1.20 \\
			1280ms & 646ms \textcolor{green}{($\downarrow$633)} & 4432ms \textcolor{green}{($\downarrow$374)} & 6563 / 10570 = 62.0\% &  1217 / 7712 = 15.7\% & 24179 / 74220 = 32.5\% & 5.81 / 5.05 & 0.08486 / 0.09876  & 1.26 \\
			\midrule
			\multicolumn{9}{c}{320ms chunk size with 114559 total chunks} \\
			\midrule
			0ms & 1015ms \textcolor{green}{($\sim$)} & 4575ms \textcolor{green}{($\sim$)} & - & - & - & 6.13 / 5.27 & 0.06007 / 0.06748 & - \\
			80ms & 965ms \textcolor{green}{($\downarrow$50)} & 4551ms \textcolor{green}{($\downarrow$24)} & 109 / 2406 = 4.5\% & 6 / 1770 = 0.3\% & 1263 / 25484 = 4.9\% & 6.13 / 5.27 & 0.06609 / 0.07526 & 0.22 \\
			160ms & 939ms \textcolor{green}{($\downarrow$76)} & 4524ms \textcolor{green}{($\downarrow$51)} & 289 / 4697 = 6.1\% & 33 / 3575 = 0.9\% & 28823 / 48779 = 5.9\% & 6.13 / 5.27 & 0.07446 / 0.07884 & 0.43 \\
			320ms & 762ms \textcolor{green}{($\downarrow$253)} & 4443ms \textcolor{green}{($\downarrow$132)} & 1795 / 7939 = 22\% & 224 / 7677 = 2.9\% & 9692 / 91571 = 10.5\% & 6.13 / 5.27 & 0.07974 / 0.08979 & 0.80 \\
			640ms & 641ms \textcolor{green}{($\downarrow$374)} & 4353ms \textcolor{green}{($\downarrow$222)} & 5750 / 10065 = 57\% & 1595 / 11493 = 13.8\% & 36893 / 137868 = 26.7\% & 6.13 / 5.27 & 0.09645 / 0.11290  & 1.20 \\
			1280ms & 621ms \textcolor{green}{($\downarrow$394)} & 4290ms \textcolor{green}{($\downarrow$285)} & 6509 / 10268 = 63.3\% & 2052 / 11767 = 17.4\% & 44869 / 144402 = 31.0\% & 6.13 / 5.27 & 0.13690 / 0.15990 & 1.26 \\
			\midrule
			\multicolumn{9}{c}{160ms chunk size with 225482 total chunks} \\
			\midrule
			0ms & 971ms \textcolor{green}{($\sim$)} & 4423ms \textcolor{green}{($\sim$)} & - & - & - & 6.35 / 5.39 & 0.09616 /  0.10590 & - \\
			80ms & 889ms \textcolor{green}{($\downarrow$82)} & 4428ms \textcolor{red}{($\uparrow$5)} & 231 / 3718 = 6.2\% & 3 / 835 = 0.3\% & 1655 / 51827 = 3.1\% & 6.35 / 5.39 & 0.10830 / 0.12070 & 0.23 \\
			160ms & 826ms \textcolor{green}{($\downarrow$145)} & 4446ms \textcolor{red}{($\uparrow$23)} & 659 / 6552 = 10.0\% & 53 / 4294 = 1.2\% & 4995 / 99480 = 5.0\% & 6.35 / 5.39 & 0.11180 / 0.12530 & 0.44 \\
			320ms & 700ms \textcolor{green}{($\downarrow$271)} & 4388ms \textcolor{green}{($\downarrow$35)} & 2150 / 7549 = 28.4\% & 276 / 7785 = 3.5\% & 18433 / 182513 = 10.0\% & 6.35 / 5.39 & 0.13040 / 0.14710 & 0.81 \\
			640ms & 574ms \textcolor{green}{($\downarrow$397)} & 4271ms \textcolor{green}{($\downarrow$152)} & 5894 / 8527 = 69\% & 2104 / 12304 = 17.1\% & 73053 / 275551 = 26.5\% & 6.35 / 5.39 & 0.16700 / 0.19220 & 1.22 \\
			1280ms & 549ms \textcolor{green}{($\downarrow$422)} & 4234ms \textcolor{green}{($\downarrow$189)} & 6761 / 8918 = 75.8\% & 2612 / 12544 = 20.8\% & 89647 / 289123 = 31.0\% & 6.35 / 5.39 & 0.24220 / 0.28250 & 1.28 \\
			\midrule
			\multicolumn{9}{c}{Librispeech (test\_clean), 283993 total characters, 2620 total sentences} \\
			\midrule
			\multicolumn{9}{c}{640ms chunk size with 31381 total chunks} \\
			\midrule
			0ms & 1136ms \textcolor{green}{($\sim$)} & 7328ms \textcolor{green}{($\sim$)} & - & - & - & 4.41 / 3.80 & 0.04826 / 0.05644 & - \\
			80ms & 1038ms \textcolor{green}{($\downarrow$98)} & 7280ms \textcolor{green}{($\downarrow$48)} & 60 / 4501 = 1.3\% & 35 / 1607 = 2.1\% & 459 / 35507 = 1.2\% & 4.41 / 3.80 & 0.05184 / 0.06111 & 1.13 \\
			160ms & 935ms \textcolor{green}{($\downarrow$201)} & 7235ms \textcolor{green}{($\downarrow$93)} & 146 / 8344 = 1.7\% & 49 / 3040 = 1.6\% & 1112 / 68465 = 1.6\% & 4.41 / 3.80 & 0.05543 / 0.06435 & 2.18 \\
			320ms & 761ms \textcolor{green}{($\downarrow$375)} & 7149ms \textcolor{green}{($\downarrow$179)} & 812 / 13916 = 5.8\% & 230 / 5929 = 3.8\% & 5667 / 123304 = 4.5\% & 4.41 / 3.80 & 0.05951 / 0.06979 & 3.93 \\
			640ms & 662ms \textcolor{green}{($\downarrow$474)} & 7098ms \textcolor{green}{($\downarrow$230)} & 2552 / 17570 = 14.5\% & 577 / 8002 = 7.2\% & 16743 / 159472 = 10.4\% & 4.41 / 3.80 & 0.07006 / 0.08295  & 5.08 \\
			1280ms & 658ms \textcolor{green}{($\downarrow$478)} & 7091ms \textcolor{green}{($\downarrow$237)} & 2522 / 17696 = 14.2\% & 658 / 8372 = 7.8\% & 18085 / 162531 = 11.1\% & 4.41 / 3.80 & 0.09096 / 0.10720  & 5.18 \\
			\midrule
			\multicolumn{9}{c}{320ms chunk size with 61432 total chunks} \\
			\midrule
			0ms & 928ms \textcolor{green}{($\sim$)} & 7147ms \textcolor{green}{($\sim$)} & - & - & - & 4.76 / 4.04 & 0.06996 / 0.08025 & - \\
			80ms & 853ms \textcolor{green}{($\downarrow$75)} & 7128ms \textcolor{green}{($\downarrow$19)} & 67 / 5185 = 1.2\% & 71 / 2552 = 2.7\% & 1041 / 70996 = 1.4\% & 4.76 / 4.04 & 0.07476 / 0.08630 & 1.16 \\
			160ms & 789ms \textcolor{green}{($\downarrow$139)} & 7091ms \textcolor{green}{($\downarrow$56)} & 157 / 9028 = 1.7\% & 69 / 5103 = 1.3\% & 2372 / 136656 = 1.7\% & 4.76 / 4.04 & 0.08155 / 0.09210 & 2.22 \\
			320ms & 662ms \textcolor{green}{($\downarrow$266)} & 7005ms \textcolor{green}{($\downarrow$142)} & 839 / 13855 = 6.0\% & 363 / 10664 = 3.4\% & 10814 / 246692 = 4.3\% & 4.76 / 4.04 & 0.08963 / 0.10370 & 4.02 \\
			640ms & 569ms \textcolor{green}{($\downarrow$359)} & 6950ms \textcolor{green}{($\downarrow$197)} & 2361 / 15297 = 15.0\% & 977 / 13863 = 7.0\% & 29997 / 317552 = 9.4\% & 4.76 / 4.04 & 0.10890 / 0.12630 & 5.17 \\
			1280ms & 561ms \textcolor{green}{($\downarrow$367)} & 6945ms \textcolor{green}{($\downarrow$202)} & 2389 / 15241 = 15.6\% & 1135 / 14228 = 7.9\% & 32287 / 323612 = 9.9\% & 4.76 / 4.04 & 0.14990 / 0.17770 & 5.28 \\
			\midrule
			\multicolumn{9}{c}{160ms chunk size with 121531 total chunks} \\
			\midrule
			0ms & 857ms \textcolor{green}{($\sim$)} & 7043ms \textcolor{green}{($\sim$)} & - & - & - & 5.10 / 4.30 & 0.11770 / 0.12970 & - \\
			80ms & 786ms \textcolor{green}{($\downarrow$71)} & 7063ms \textcolor{red}{($\uparrow$20)} & 65 / 5395 = 1.2\% & 59 / 2345 = 2.5\% & 1462 / 140612 = 1.0\% & 5.10 / 4.30 & 0.12880 / 0.14350 & 1.16 \\
			160ms & 704ms \textcolor{green}{($\downarrow$153)} & 7048ms \textcolor{red}{($\uparrow$5)} & 135 / 10685 = 1.2\% & 84 / 5830 = 1.4\% & 3833 / 271459 = 1.4\% & 5.10 / 4.30 & 0.13480 / 0.15050 & 2.23 \\
			320ms & 579ms \textcolor{green}{($\downarrow$278)} & 6959ms \textcolor{green}{($\downarrow$84)} & 833 / 11942 = 6.9\% & 470 / 11533 = 4.0\% & 16573 / 493650 = 3.3\% & 5.10 / 4.30 & 0.14760 / 0.17060 & 4.06 \\
			640ms & 505ms \textcolor{green}{($\downarrow$352)} & 6909ms \textcolor{green}{($\downarrow$134)} & 2246 / 12438 = 18\% & 1274 / 14642 = 8.7\% & 44768 / 638700 = 7.0\% & 5.10 / 4.30 & 0.19030 / 0.22190 & 5.26 \\
			1280ms & 502ms \textcolor{green}{($\downarrow$355)} & 6903ms \textcolor{green}{($\downarrow$140)} & 2381 / 12612 = 18.8\% & 1438 / 15262 = 9.4\% & 48181 / 649938 = 7.4\% & 5.10 / 4.30 & 0.26960 / 0.31480 & 5.35 \\
			\midrule
			\multicolumn{9}{c}{Librispeech (test\_other), 274213 total characters, 2939 total sentences} \\
			\midrule
			\multicolumn{9}{c}{640ms chunk size with 31120 total chunks} \\
			\midrule
			0ms & 1209ms \textcolor{green}{($\sim$)} & 6428ms \textcolor{green}{($\sim$)} & - & - & - & 11.48 / 10.40 & 0.04826 / 0.05644 & - \\
			80ms & 1130ms \textcolor{green}{($\downarrow$79)} & 6407ms \textcolor{green}{($\downarrow$21)} & 126 / 5013 = 2.5\% & 47 / 1708 = 2.7\% & 800 / 33840 = 2.3\% & 11.48 / 10.40 & 0.05184 / 0.06111 & 1.09 \\
			160ms & 1032ms \textcolor{green}{($\downarrow$177)} & 6362ms \textcolor{green}{($\downarrow$66)} & 366 / 9226 = 3.9\% & 110 / 3243 = 3.3\% & 2342 / 66182 = 3.5\% & 11.48 / 10.40 & 0.05543 / 0.06435 & 2.13 \\
			320ms & 821ms \textcolor{green}{($\downarrow$388)} & 6252ms \textcolor{green}{($\downarrow$176)} & 1545 / 15336 = 10.0\% & 371 / 6970 = 5.3\% & 9839 / 121799 = 8.0\% & 11.48 / 10.40 & 0.05951 / 0.06979 & 3.91 \\
			640ms & 668ms \textcolor{green}{($\downarrow$541)} & 6208ms \textcolor{green}{($\downarrow$220)} & 3446 / 18363 = 18.7\% & 904 / 9195 = 9.8\% & 23456 / 158130 = 14.8\% & 11.48 / 10.40 & 0.07006 / 0.08295 & 5.08 \\
			1280ms & 665ms \textcolor{green}{($\downarrow$544)} & 6202ms \textcolor{green}{($\downarrow$226)} & 3598 / 18616 = 19.3\% & 1033 / 9549 = 10.8\% & 24264 / 160299 = 15.1\% & 11.48 / 10.40 & 0.09096 / 0.10720 & 5.15 \\
			\midrule
			\multicolumn{9}{c}{320ms chunk size with 60793 total chunks} \\
			\midrule
			0ms & 978ms \textcolor{green}{($\sim$)} & 6215ms \textcolor{green}{($\sim$)} & - & - & - & 12.19 / 11.06 & 0.06996 / 0.08025 & - \\
			80ms & 898ms \textcolor{green}{($\downarrow$80)} & 6235ms \textcolor{red}{($\uparrow$20)} & 150 / 5671 = 2.6\% & 79 / 2773 = 2.8\% & 1606 / 67888 = 2.3\% & 12.19 / 11.06 & 0.07476 / 0.08630 & 1.12 \\
			160ms & 840ms \textcolor{green}{($\downarrow$138)} & 6194ms \textcolor{green}{($\downarrow$21)} & 378 / 10017 = 3.7\% & 130 / 5570 = 2.3\% & 4445 / 131505 = 3.3\% & 12.19 / 11.06 & 0.08155 / 0.09210 & 2.16 \\
			320ms & 716ms \textcolor{green}{($\downarrow$262)} & 6108ms \textcolor{green}{($\downarrow$107)} & 1526 / 15766 = 9.6\% & 595 / 12212 = 4.8\% & 17909 / 241994 = 7.4\% & 12.19 / 11.06 & 0.08963 / 0.10370 & 3.98 \\
			640ms & 613ms \textcolor{green}{($\downarrow$365)} & 6052ms \textcolor{green}{($\downarrow$163)} & 3339 / 17106 = 19.5\% & 1578 / 15498 = 10.1\% & 41238 / 312230 = 13.2\% & 12.19 / 11.06 & 0.10890 / 0.12630 & 5.13 \\
			1280ms & 611ms \textcolor{green}{($\downarrow$367)} & 6051ms \textcolor{green}{($\downarrow$164)} & 3517 / 17336 = 20.2\% & 1699 / 15953 = 10.6\% & 42756 / 316521 = 13.5\% & 12.19 / 11.06 & 0.14990 / 0.17770 & 5.21 \\
			\midrule
			\multicolumn{9}{c}{160ms chunk size with 120144 total chunks} \\
			\midrule
			0ms & 909ms \textcolor{green}{($\sim$)} & 6095ms \textcolor{green}{($\sim$)} & - & - & - & 13.14 / 11.85 & 0.11770 / 0.12970 & - \\
			80ms & 835ms \textcolor{green}{($\downarrow$74)} & 6124ms \textcolor{red}{($\uparrow$29)} & 114 / 6222 = 1.8\% & 70 / 3211 = 2.1\% & 2192 / 134209 = 1.6\% & 13.14 / 11.85 & 0.12880 / 0.14350 & 1.11 \\
			160ms & 758ms \textcolor{green}{($\downarrow$151)} & 6108ms \textcolor{red}{($\uparrow$13)} & 363 / 12015 = 3.0\% & 126 / 7313 = 1.7\% & 6349 / 262655 = 2.4\% & 13.14 / 11.85 & 0.13480 / 0.15050 & 2.19 \\
			320ms & 629ms \textcolor{green}{($\downarrow$280)} & 6045ms \textcolor{green}{($\downarrow$50)} & 1467 / 13593 = 10.7\% & 931 / 14206 = 6.5\% & 26746 / 487285 = 5.4\% & 13.14 / 11.85 & 0.14760 / 0.17060 & 4.06 \\
			640ms & 552ms \textcolor{green}{($\downarrow$357)} & 5990ms \textcolor{green}{($\downarrow$105)} & 3169 / 14189 = 22.3\% & 2239 / 17967 = 12.4\% & 60588 / 628283 = 9.6\% & 13.14 / 11.85 & 0.19030 / 0.22190 & 5.23 \\
			1280ms & 548ms \textcolor{green}{($\downarrow$361)} & 5982ms \textcolor{green}{($\downarrow$113)} & 3319 / 14385 = 23.0\% & 2435 / 18567 = 13.1\% & 62239 / 636094 = 9.7\% & 13.14 / 11.85 & 0.26960 / 0.31480 & 5.29 \\
			\bottomrule[1.5pt]
		\end{tabular}
	}
	\label{tab:results}
\end{table*}

We further provide a concrete example to compare ZeroPrompt with other methods in Figure 2. It should be noted that it's reasonable for the predictions from the first ZeroPrompt chunk to be inaccurate due to the lack of contextual information. However, this is not a significant issue since most of the errors are homophones of the correct counterparts, i.e., (``\begin{CJK}{UTF8}{gbsn}\textbf{\textcolor{lightgray}{剩}}, sheng in English\end{CJK}'') vs. (``\begin{CJK}{UTF8}{gbsn}\textbf{\textcolor{black}{甚}}, shen in English\end{CJK}'') in this example. Additionally, these errors will be quickly corrected by the Prompt-and-Refine strategy, as demonstrated in Figure 3.

\section{Experiments}
To demonstrate the effectiveness of our proposed ZeroPrompt, we carry out our experiments on the open-source Chinese Mandarin speech corpus Aishell-1~\cite{aishell-1} and English speech corpus Librispeech~\cite{librispeech}. ZeroPrompt is a training-free method and can be directly applied to a well-trained chunk-based ASR model. To ensure the reproducibility of experiments, we used checkpoints downloaded from the official WeNet~\cite{wenet} website for all of our baseline models and keep the exact same settings as in open-sourced Aishell-1 and Librispeech recipes.

\subsection{Metrics}
Besides \textbf{Token Display Time} (TDT, TDT-F for First token and TDT-L for Last token), \textbf{Word Error Rate} (WER) and \textbf{Real Time Factor} (RTF), we propose several additional metrics to better analyze the effectiveness of ZeroPrompt. Specifically, we introduce two new metrics that are designed for ZeroPrompt:
\begin{itemize}
	\item \textbf{Prompts Error Rate} (PER, PER-F for First chunk, PER-L for Last chunk and PER-A for All chunks): PER is calculated by dividing Prompt Errors (PE) by the Number of Prompts (NP). NP represents the number of future characters decoded from the ZeroPrompt chunk, while PE denotes the number of errors that occur among those characters.
	\item \textbf{Prompts Per Chunk} (PPC): The PPC is obtained by dividing the total number of prompts by the total number of chunks. This metric provides insight into the average number of future characters prompted per chunk.
\end{itemize}

\subsection{Main Results}
We present the main results of ZeroPrompt in Table 1, from which 5 conclusions can be deduced:
\begin{itemize}
	\item A larger ZeroPrompt length generally results in lower Token Display Time (TDT) for all languages and chunk sizes. However, when the length exceeds a certain threshold (i.e., greater than 640ms), there is a latency ceiling imposed by both the chunk size (TDT cannot be smaller than chunk size due to the required data collecting time) and the leading silence (the ASR model cannot prompt tokens if both the current chunk and the ZeroPrompt chunk contain only silences or zeros).
	\item A larger ZeroPrompt length also results in a higher PER, but this is not a significant problem because they can be rapidly corrected using our Prompt-and-Refine strategy, which is described in Section 2 and illustrated in Figure 3.
	\item The closer a chunk is to the end of a sentence, the more accurate the prompts are. It is clear that PER-L is much better than PER-F, which is reasonable because the first tokens often lack context information while the last tokens have richer context.
	\item Thanks to the autoregressive attention mask (Figure 1-(b)) and the Prompt-and-Refine strategy (Figure 3), the WER for the final result remained unchanged. However, we observed a slight increase in RTF due to the increased input length. It's worth noting that, if compared within a similar RTF, \textit{\textcolor{green}{[640ms chunk size \& 640ms ZeroPrompt, Aishell-1, RTF 0.06447/0.07425]}} significantly outperforms \textit{\textcolor{red}{[320ms chunk size \& 0ms ZeroPrompt, Aishell-1, RTF 0.06007/0.06748]}} in TDT-F (646ms v.s. 1015ms), TDT-L (4472ms v.s. 4575ms) and WER (5.81/5.05 v.s. 6.13/5.27). This is mainly because the 640ms chunk size provides more context information than the 320ms chunk size, and the 640ms ZeroPrompt greatly reduces latency compared to the 0ms baseline. Moreover, to offer users greater flexibility in balancing latency (TDT \& PPC) and RTF, we further discuss a solution in Section 3.3.
	\item It appears that PPC only correlates with ZeroPrompt length, as different chunk sizes result in similar PPC values.
\end{itemize}

Overall, based on the results from Aishell-1, we can conclude that ZeroPrompt provides the best trade-off between latency (TDT \& PPC) and WER, both theoretically and experimentally. It achieves a reduction of 350 $\sim$ 700ms in TDT-F and 100 $\sim$ 400ms in TDT-L, while keeping WER unchanged. This conclusion is further supported by the results from Librispeech, which demonstrate that ZeroPrompt generalizes well to any dataset without requiring any extra effort.

\subsection{Solution to balance latency-RTF Trade-off}
As described in Section 3.2, although the latency-WER trade-off has been solved, there is also a trade-off between latency (TDT \& PPC) and RTF. In this section, we present a solution, called \textbf{Intermediate ZeroPrompt}, to better balance latency and RTF. Specifically, we feed the ZeroPrompt chunk starting from different encoder layers to achieve different computation costs. From Table 2, it can be observed that one can simply change the start layer to meet the desired latency and RTF requirements.

\begin{table}[!htp]
	\centering
        \vspace{-5pt}
	\caption{Results of Intermediate ZeroPrompt [640ms chunk size \& 640ms ZeroPrompt, Aishell-1]. 0 means we feed ZeroPrompt chunk to the first encoder layer and this is the default ZeroPrompt method used in Table 1. -1 means baseline without ZeroPrompt.}
        \vspace{-10pt}
	\scalebox{0.8}{
		\begin{tabular}{ccccc}
			\toprule[1.5pt]
			StartLayer & TDT-F & TDT-L & PPC & RTF \\
			\midrule
			0 & 646ms & 4472ms & 1.20 & 0.06447 / 0.07425 \\
			4 & 649ms & 4477ms & 1.20 & 0.05779 / 0.06906 \\
			6 & 778ms & 4562ms & 0.97 & 0.05444 / 0.06596 \\
			8 & 1099ms & 4652ms & 0.70 & 0.05186 / 0.06273 \\
			11 & 1149ms & 4815ms & 0.34 & 0.04734 / 0.05858 \\
			\midrule
			-1 & 1279ms & 4806ms & - & 0.04351 / 0.05063 \\
			\bottomrule[1.5pt]
		\end{tabular}
	}
        \vspace{-10pt}
	\label{tab:trade-off}
\end{table}

\subsection{Error Analysis}
Lastly, we provide error analysis on \textit{[640ms chunk size \& 1280ms ZeroPrompt, Aishell-1]} as this configuration achieves the worst PER and the best PPC. We find that errors can be categried into two types:
\begin{itemize}
	\item \textbf{Homophonic tokens}, typically occur at the beginning of prompts. This is reasonable  because the current chunk may only contain partial pronunciations of the character, and ZeroPrompt forces the model to emit a complete character based on these partial pronunciations thus leading to homophone errors.
	\item \textbf{Semantically continuous but phonetically mismatched tokens}, typically occur at the end of a \textbf{very long} prompt. The trailing part of ZeroPrompt chunk contains no partial pronunciation, therefore the prediction of trailing prompts solely depends on the history context without any acoustic hints, like a Masked LM, this further validate our conjecture that streaming ascoutic encoders are zero-shot Masked LMs.
\end{itemize}

\bibliographystyle{IEEEtran}
\bibliography{mybib}

\begin{thebibliography}{10}
\providecommand{\url}[1]{#1}
\csname url@samestyle\endcsname
\providecommand{\newblock}{\relax}
\providecommand{\bibinfo}[2]{#2}
\providecommand{\BIBentrySTDinterwordspacing}{\spaceskip=0pt\relax}
\providecommand{\BIBentryALTinterwordstretchfactor}{4}
\providecommand{\BIBentryALTinterwordspacing}{\spaceskip=\fontdimen2\font plus
\BIBentryALTinterwordstretchfactor\fontdimen3\font minus
  \fontdimen4\font\relax}
\providecommand{\BIBforeignlanguage}[2]{{%
\expandafter\ifx\csname l@#1\endcsname\relax
\typeout{** WARNING: IEEEtran.bst: No hyphenation pattern has been}%
\typeout{** loaded for the language `#1'. Using the pattern for}%
\typeout{** the default language instead.}%
\else
\language=\csname l@#1\endcsname
\fi
#2}}
\providecommand{\BIBdecl}{\relax}
\BIBdecl

\bibitem{ctc}
\BIBentryALTinterwordspacing
A.~Graves, S.~Fern{\'{a}}ndez, F.~J. Gomez, and J.~Schmidhuber, ``Connectionist
  temporal classification: labelling unsegmented sequence data with recurrent
  neural networks,'' in \emph{Machine Learning, Proceedings of the Twenty-Third
  International Conference {(ICML} 2006), Pittsburgh, Pennsylvania, USA, June
  25-29, 2006}, ser. {ACM} International Conference Proceeding Series, W.~W.
  Cohen and A.~W. Moore, Eds., vol. 148.\hskip 1em plus 0.5em minus 0.4em\relax
  {ACM}, 2006, pp. 369--376. [Online]. Available:
  \url{https://doi.org/10.1145/1143844.1143891}
\BIBentrySTDinterwordspacing

\bibitem{rnnt}
\BIBentryALTinterwordspacing
A.~Graves, ``Sequence transduction with recurrent neural networks,''
  \emph{CoRR}, vol. abs/1211.3711, 2012. [Online]. Available:
  \url{http://arxiv.org/abs/1211.3711}
\BIBentrySTDinterwordspacing

\bibitem{speech-transformer}
\BIBentryALTinterwordspacing
L.~Dong, S.~Xu, and B.~Xu, ``Speech-transformer: {A} no-recurrence
  sequence-to-sequence model for speech recognition,'' in \emph{2018 {IEEE}
  International Conference on Acoustics, Speech and Signal Processing, {ICASSP}
  2018, Calgary, AB, Canada, April 15-20, 2018}.\hskip 1em plus 0.5em minus
  0.4em\relax {IEEE}, 2018, pp. 5884--5888. [Online]. Available:
  \url{https://doi.org/10.1109/ICASSP.2018.8462506}
\BIBentrySTDinterwordspacing

\bibitem{dual}
\BIBentryALTinterwordspacing
J.~Yu, W.~Han, A.~Gulati, C.~Chiu, B.~Li, T.~N. Sainath, Y.~Wu, and R.~Pang,
  ``Dual-mode {ASR:} unify and improve streaming {ASR} with full-context
  modeling,'' in \emph{9th International Conference on Learning
  Representations, {ICLR} 2021, Virtual Event, Austria, May 3-7, 2021}.\hskip
  1em plus 0.5em minus 0.4em\relax OpenReview.net, 2021. [Online]. Available:
  \url{https://openreview.net/forum?id=Pz\_dcqfcKW8}
\BIBentrySTDinterwordspacing

\bibitem{saa}
\BIBentryALTinterwordspacing
L.~Dong, F.~Wang, and B.~Xu, ``Self-attention aligner: {A} latency-control
  end-to-end model for {ASR} using self-attention network and chunk-hopping,''
  in \emph{{IEEE} International Conference on Acoustics, Speech and Signal
  Processing, {ICASSP} 2019, Brighton, United Kingdom, May 12-17, 2019}.\hskip
  1em plus 0.5em minus 0.4em\relax {IEEE}, 2019, pp. 5656--5660. [Online].
  Available: \url{https://doi.org/10.1109/ICASSP.2019.8682954}
\BIBentrySTDinterwordspacing

\bibitem{u2}
\BIBentryALTinterwordspacing
B.~Zhang, D.~Wu, Z.~Yao, X.~Wang, F.~Yu, C.~Yang, L.~Guo, Y.~Hu, L.~Xie, and
  X.~Lei, ``Unified streaming and non-streaming two-pass end-to-end model for
  speech recognition,'' \emph{CoRR}, vol. abs/2012.05481, 2020. [Online].
  Available: \url{https://arxiv.org/abs/2012.05481}
\BIBentrySTDinterwordspacing

\bibitem{blstm}
M.~Schuster and K.~Paliwal, ``Bidirectional recurrent neural networks,''
  \emph{IEEE Transactions on Signal Processing}, vol.~45, no.~11, pp.
  2673--2681, 1997.

\bibitem{transformer}
\BIBentryALTinterwordspacing
A.~Vaswani, N.~Shazeer, N.~Parmar, J.~Uszkoreit, L.~Jones, A.~N. Gomez,
  L.~Kaiser, and I.~Polosukhin, ``Attention is all you need,'' in
  \emph{Advances in Neural Information Processing Systems 30: Annual Conference
  on Neural Information Processing Systems 2017, December 4-9, 2017, Long
  Beach, CA, {USA}}, I.~Guyon, U.~von Luxburg, S.~Bengio, H.~M. Wallach,
  R.~Fergus, S.~V.~N. Vishwanathan, and R.~Garnett, Eds., 2017, pp. 5998--6008.
  [Online]. Available:
  \url{https://proceedings.neurips.cc/paper/2017/hash/3f5ee243547dee91fbd053c1c4a845aa-Abstract.html}
\BIBentrySTDinterwordspacing

\bibitem{upl}
\BIBentryALTinterwordspacing
Y.~Shangguan, R.~Prabhavalkar, H.~Su, J.~Mahadeokar, Y.~Shi, J.~Zhou, C.~Wu,
  D.~Le, O.~Kalinli, C.~Fuegen, and M.~L. Seltzer, ``Dissecting user-perceived
  latency of on-device {E2E} speech recognition,'' in \emph{Interspeech 2021,
  22nd Annual Conference of the International Speech Communication Association,
  Brno, Czechia, 30 August - 3 September 2021}, H.~Hermansky,
  H.~Cernock{\'{y}}, L.~Burget, L.~Lamel, O.~Scharenborg, and
  P.~Motl{\'{\i}}cek, Eds.\hskip 1em plus 0.5em minus 0.4em\relax {ISCA}, 2021,
  pp. 4553--4557. [Online]. Available:
  \url{https://doi.org/10.21437/Interspeech.2021-1887}
\BIBentrySTDinterwordspacing

\bibitem{fastemit}
\BIBentryALTinterwordspacing
J.~Yu, C.~Chiu, B.~Li, S.~Chang, T.~N. Sainath, Y.~He, A.~Narayanan, W.~Han,
  A.~Gulati, Y.~Wu, and R.~Pang, ``Fastemit: Low-latency streaming {ASR} with
  sequence-level emission regularization,'' in \emph{{IEEE} International
  Conference on Acoustics, Speech and Signal Processing, {ICASSP} 2021,
  Toronto, ON, Canada, June 6-11, 2021}.\hskip 1em plus 0.5em minus 0.4em\relax
  {IEEE}, 2021, pp. 6004--6008. [Online]. Available:
  \url{https://doi.org/10.1109/ICASSP39728.2021.9413803}
\BIBentrySTDinterwordspacing

\bibitem{peak-first-ctc}
\BIBentryALTinterwordspacing
Z.~Tian, H.~Xiang, M.~Li, F.~Lin, K.~Ding, and G.~Wan, ``Peak-first {CTC:}
  reducing the peak latency of {CTC} models by applying peak-first
  regularization,'' \emph{CoRR}, vol. abs/2211.03284, 2022. [Online].
  Available: \url{https://doi.org/10.48550/arXiv.2211.03284}
\BIBentrySTDinterwordspacing

\bibitem{trimtail}
\BIBentryALTinterwordspacing
X.~Song, D.~Wu, Z.~Wu, B.~Zhang, Y.~Zhang, Z.~Peng, W.~Li, F.~Pan, and C.~Zhu,
  ``Trimtail: Low-latency streaming {ASR} with simple but effective
  spectrogram-level length penalty,'' \emph{CoRR}, vol. abs/2211.00522, 2022.
  [Online]. Available: \url{https://doi.org/10.48550/arXiv.2211.00522}
\BIBentrySTDinterwordspacing

\bibitem{lookahead1}
\BIBentryALTinterwordspacing
D.~Povey, H.~Hadian, P.~Ghahremani, K.~Li, and S.~Khudanpur, ``A
  time-restricted self-attention layer for {ASR},'' in \emph{2018 {IEEE}
  International Conference on Acoustics, Speech and Signal Processing, {ICASSP}
  2018, Calgary, AB, Canada, April 15-20, 2018}.\hskip 1em plus 0.5em minus
  0.4em\relax {IEEE}, 2018, pp. 5874--5878. [Online]. Available:
  \url{https://doi.org/10.1109/ICASSP.2018.8462497}
\BIBentrySTDinterwordspacing

\bibitem{lookahead2}
C.~Wu, Y.~Wang, Y.~Shi, C.-F. Yeh, and F.~Zhang, ``{Streaming Transformer-Based
  Acoustic Models Using Self-Attention with Augmented Memory},'' in \emph{Proc.
  Interspeech 2020}, 2020, pp. 2132--2136.

\bibitem{adapter}
\BIBentryALTinterwordspacing
K.~Deng and P.~C. Woodland, ``Adaptable end-to-end {ASR} models using
  replaceable internal lms and residual softmax,'' \emph{CoRR}, vol.
  abs/2302.08579, 2023. [Online]. Available:
  \url{https://doi.org/10.48550/arXiv.2302.08579}
\BIBentrySTDinterwordspacing

\bibitem{bert}
\BIBentryALTinterwordspacing
J.~Devlin, M.~Chang, K.~Lee, and K.~Toutanova, ``{BERT:} pre-training of deep
  bidirectional transformers for language understanding,'' in \emph{Proceedings
  of the 2019 Conference of the North American Chapter of the Association for
  Computational Linguistics: Human Language Technologies, {NAACL-HLT} 2019,
  Minneapolis, MN, USA, June 2-7, 2019, Volume 1 (Long and Short Papers)},
  J.~Burstein, C.~Doran, and T.~Solorio, Eds.\hskip 1em plus 0.5em minus
  0.4em\relax Association for Computational Linguistics, 2019, pp. 4171--4186.
  [Online]. Available: \url{https://doi.org/10.18653/v1/n19-1423}
\BIBentrySTDinterwordspacing

\bibitem{revision}
\BIBentryALTinterwordspacing
Z.~Li, H.~Miao, K.~Deng, G.~Cheng, S.~Tian, T.~Li, and Y.~Yan, ``Improving
  streaming end-to-end {ASR} on transformer-based causal models with encoder
  states revision strategies,'' in \emph{Interspeech 2022, 23rd Annual
  Conference of the International Speech Communication Association, Incheon,
  Korea, 18-22 September 2022}, H.~Ko and J.~H.~L. Hansen, Eds.\hskip 1em plus
  0.5em minus 0.4em\relax {ISCA}, 2022, pp. 1671--1675. [Online]. Available:
  \url{https://doi.org/10.21437/Interspeech.2022-707}
\BIBentrySTDinterwordspacing

\bibitem{cuside}
\BIBentryALTinterwordspacing
K.~An, H.~Zheng, Z.~Ou, H.~Xiang, K.~Ding, and G.~Wan, ``{CUSIDE:} chunking,
  simulating future context and decoding for streaming {ASR},'' in
  \emph{Interspeech 2022, 23rd Annual Conference of the International Speech
  Communication Association, Incheon, Korea, 18-22 September 2022}, H.~Ko and
  J.~H.~L. Hansen, Eds.\hskip 1em plus 0.5em minus 0.4em\relax {ISCA}, 2022,
  pp. 2103--2107. [Online]. Available:
  \url{https://doi.org/10.21437/Interspeech.2022-11214}
\BIBentrySTDinterwordspacing

\bibitem{apc}
\BIBentryALTinterwordspacing
Y.~Chung and J.~R. Glass, ``Generative pre-training for speech with
  autoregressive predictive coding,'' in \emph{2020 {IEEE} International
  Conference on Acoustics, Speech and Signal Processing, {ICASSP} 2020,
  Barcelona, Spain, May 4-8, 2020}.\hskip 1em plus 0.5em minus 0.4em\relax
  {IEEE}, 2020, pp. 3497--3501. [Online]. Available:
  \url{https://doi.org/10.1109/ICASSP40776.2020.9054438}
\BIBentrySTDinterwordspacing

\bibitem{aishell-1}
\BIBentryALTinterwordspacing
H.~Bu, J.~Du, X.~Na, B.~Wu, and H.~Zheng, ``{AISHELL-1:} an open-source
  mandarin speech corpus and a speech recognition baseline,'' in \emph{20th
  Conference of the Oriental Chapter of the International Coordinating
  Committee on Speech Databases and Speech {I/O} Systems and Assessment,
  {O-COCOSDA} 2017, Seoul, South Korea, November 1-3, 2017}.\hskip 1em plus
  0.5em minus 0.4em\relax {IEEE}, 2017, pp. 1--5. [Online]. Available:
  \url{https://doi.org/10.1109/ICSDA.2017.8384449}
\BIBentrySTDinterwordspacing

\bibitem{librispeech}
\BIBentryALTinterwordspacing
V.~Panayotov, G.~Chen, D.~Povey, and S.~Khudanpur, ``Librispeech: An {ASR}
  corpus based on public domain audio books,'' in \emph{2015 {IEEE}
  International Conference on Acoustics, Speech and Signal Processing, {ICASSP}
  2015, South Brisbane, Queensland, Australia, April 19-24, 2015}.\hskip 1em
  plus 0.5em minus 0.4em\relax {IEEE}, 2015, pp. 5206--5210. [Online].
  Available: \url{https://doi.org/10.1109/ICASSP.2015.7178964}
\BIBentrySTDinterwordspacing

\bibitem{wenet}
\BIBentryALTinterwordspacing
B.~Zhang, D.~Wu, Z.~Peng, X.~Song, Z.~Yao, H.~Lv, L.~Xie, C.~Yang, F.~Pan, and
  J.~Niu, ``Wenet 2.0: More productive end-to-end speech recognition toolkit,''
  in \emph{Interspeech 2022, 23rd Annual Conference of the International Speech
  Communication Association, Incheon, Korea, 18-22 September 2022}, H.~Ko and
  J.~H.~L. Hansen, Eds.\hskip 1em plus 0.5em minus 0.4em\relax {ISCA}, 2022,
  pp. 1661--1665. [Online]. Available:
  \url{https://doi.org/10.21437/Interspeech.2022-483}
\BIBentrySTDinterwordspacing

\end{thebibliography}

\end{document}